\documentclass[pre,a4paper,oneside,twocolumn]{revtex4-1}

\usepackage[english]{babel}
\usepackage[latin1]{inputenc}
\usepackage{graphicx}  
\usepackage{latexsym}  
\usepackage{amssymb,amsmath,amsthm,amsfonts,braket,dsfont}	

\begin{document}

\title{Scaling Function, Universality and  Analytical Solutions of Generalized One-Species Population Dynamics Models}

\author{\firstname{Alexandre} Souto \surname{Martinez}}
\email{asmartinez@ffclrp.usp.br}
\affiliation{Departamento de F\'isica e Matem\'atica (DFM) \\
             Faculdade de Filosofia, Ci\^encias e Letras de Ribeir\~ao Preto (FFCLRP), \\
             Universidade de S\~ao Paulo (USP) \\ 
             Avenida Bandeirantes, 3900 \\ 
             14040-901, Ribeir\~ao Preto, S\~ao Paulo, Brazil.}
\affiliation{National Institute of Science and Technology in Complex Systems (LNCT-SC)}

\author{\firstname{Brenno} Caetano Troca \surname{Cabella}}
\email{brenno@usp.br}
\affiliation{Departamento de F\'isica e Matem\'atica (DFM) \\
             Faculdade de Filosofia, Ci\^encias e Letras de Ribeir\~ao Preto (FFCLRP), \\
             Universidade de S\~ao Paulo (USP) \\ 
             Avenida Bandeirantes, 3900 \\ 
             14040-901, Ribeir\~ao Preto, S\~ao Paulo, Brazil.}
\affiliation{National Institute of Science and Technology in Complex Systems (LNCT-SC)}

\author{\firstname{Fabiano}  \surname{Ribeiro}}
\email{flribeiro@dex.ufla.br}
\affiliation{Departamento de Ci\^encias Exatas (DEX), \\
                Universidade Federal de Lavras (UFLA) \\
	           Caixa Postal 3037 \\
		     37200-000 Lavras, MG, Brazil.}
\affiliation{National Institute of Science and Technology in Complex Systems (LNCT-SC)}

\date{\today}

\begin{abstract}
We consider several one-species population dynamics model with finite and infinite carrying capacity, time dependent growth and effort rates and solve them analytically.
We show that defining suitable scaling functions for a given time, one is able to demonstrate that their ratio with respect to its initial value is  universal.
This ratio is independent from the initial condition and from the model parameters. 
Although the effort rate does not break the model universality it produces a transition between the species extinction and survival. 
A general formula is furnished to obtain the scaling functions.
\end{abstract}

\keywords{Complex Systems,  Population dynamics (ecology), Nonlinear dynamics}
\pacs{89.75.-k, 87.23.-n, 87.23.Cc, 05.45.-a}

\maketitle



\section{Introduction}
\label{introduction}
The population growth problem is one of the most important of the scientific knowledge. 
It is crucial not only to describe ecological systems like bacterial or  virus populations,  but also to understand  economic behaviors and human population growth~\cite{boyce_diprima,murray,Keshet,nowak,begon,sigmund}.  

The simplest way to deal with population growth is to consider that their individuals do not interact with external ones. 
This is represented by the so-called one-species population dynamics models. 
These models quantify the size (number of individuals) $N(t) \ge 0$ of a given population, at a certain time $t$, given its: initial size $N_0 \equiv N(0) > 0$; growth rate $\kappa > 0$ and the environmental carrying capacity $K = N(\infty) > 0$.  
The environmental carrying capacity takes into account all possible interactions among individuals, species and resources into a single parameter. 
If one assumes that the population lives in an environment with unlimited resources (infinite carrying capacity), the population grows  exponentially. 
The \textit{Malthus} model is an example that generates this behavior.
However, for finite carrying capacity, the growth of individual organisms~\cite{laird65}, tumours~\cite{bajzer96} and other biological systems~\cite{zwietering90} are well fitted by sigmoid curves~\cite{boyce_diprima,murray,Keshet} that can be obtained from the \textit{Gompertz} or \textit{Verhulst} model, for instance.

The generalization of the well known growth models, by the \textit{von Foerster et al.}~\cite{vonfoerster} or \textit{Richards}~\cite{richards_1959} models, introduces additional parameters to the former models, making them more complete and accurate to fit experimental data. 
In this attempt to find a suitable complete growth model, the one-parameter generalizations of the logarithm and exponential functions play a central role~\cite{tsallis_1988,tsallis_qm,arruda_2008,martinez,martinez2}. 
They allow us to easily retrieve particular cases without needing to calculate limits and permit convenient algebraic tricks to handle the expressions. 

From the analytical solutions obtained from the one-species growth models that we address, we call attention to the following items. 
For each model, the population evolution is proportional to the inverse of the growth rate. 
This gives rise to a dimensionless characteristic time and can be defined as the system independent variable $\tau$. 
Furthermore, we define a scaling function that depends on the solution of each model. 
The dependent variable $y$ is the ratio between the scaling function, at a given time $\tau$, with respect to its initial value. 
We show that using these variables, the models are independent from the initial conditions and parameters demonstrating a universal Malthusian behavior.
This universality occurs for a finite carrying capacity models and even in the \textit{Tsoularis-Wallace} model, where no closed analytical solution is found. 

Finally, we stress that when a constant effort rate is considered, the universality is preserved and the steady state (asymptotic) solution can be interpreted as an order parameter.  
A transition between extinction and survival phases are separated by a critical value, which depends on the effort rate. 
This transition does not  occur for the Gompertz model.  
This picture is not altered when time dependent effort rates is considered in the Richards' model, nor when a time-dependent growth rate is considered. 
Although a closed analytical solution is not found, we are able to show the universality of the Tsoularis-Wallace model with constant effort rate.

This manuscript is structured as follows. 
In Sec.~\ref{sec:generalized_logarithm_exponential}, we briefly review the one-parameter generalization of the logarithm and exponential functions and present some of its properties used along the study. 
In Sec.~\ref{sec:mathus_verhulst_gompertz}, we introduce the main one-species population dynamics (growth) models. 
We define the scaling function and show that all these models can be written as the universal Malthus (exponential) model.
In Sec.~\ref{sec:generalized_model_effort_rate}, we consider the insertion or removal of individuals through a constant effort rate. 
We show that this quantity does not affect the universality of the models. 
Moreover, it induces an extinction-survival transition at a well determined value.
Next, we show that time dependent effort rate and time dependent growth rate do not affect universality. 
In Sec.~\ref{conclusion}, we present a formula to obtain the scaling functions and our final remarks. 

\section{Generalized Logarithmic and Exponential Functions}
\label{sec:generalized_logarithm_exponential}

In the following, we introduce the one-parameter generalization of the logarithmic exponential function and present some of their main used properties.

The $\tilde{q}$-\textit{logarithm function} is defined as: 
\begin{equation}
\ln_{\tilde{q}}(x) = \frac{x^{\tilde{q}} -1  }{\tilde{q}} = \int_1^x \frac{dt}{t^{1-\tilde{q}}}  \; .
\label{eq-ln-q}
\end{equation}
This one-parameter generalization of the natural logarithm function, which is retrieved for $\tilde{q}\to 0$, has been introduced in the context of non-extensive statistical mechanics~\cite{tsallis_1988,tsallis_qm} and is defined as the value of the area underneath the non-symmetric hyperbole,  $f_{\tilde q}(t)=1/t^{1-\tilde q}$, in the interval $t \in [1,x]$~\cite{arruda_2008}. 
Note that  in Eq.~(\ref{eq-ln-q}), $\ln_{\tilde{q}}(x)$ is not ``logarithm $x$ in the base $\tilde{q}$''. 
For $\tilde{q} < 0$, $\ln_{\tilde{q}}(\infty)=-1/\tilde{q}$; for $\tilde{q} > 0$, $\ln_{\tilde{q}}(0)=-1/\tilde{q}$; for all $\tilde{q}$, $\ln_{\tilde{q}}(1)=0$; $\ln_{\tilde{q}}(x^{-1}) = - \ln_{-\tilde{q}}(x)$;  $d \ln_{\tilde{q}}(x)/dx = x^{\tilde{q}-1}$. 

The inverse of the $\tilde{q}$-\textit{logarithm function}  is the $\tilde{q}$-\textit{exponential function}
\begin{equation}
e_{\tilde{q}}(x) =   
\left\{ \begin{array}{ll}
\lim_{\tilde{q}^{'} \to \tilde{q}}   (1+ \tilde{q}^{'} x)^{ \frac{1}{ \tilde{q}^{'}} } & ,\textrm{ if $\tilde{q}x > -1$} \\
0 & ,  \textrm{ otherwise} 
\end{array} \right. \; ,
\label{def-eq}
\end{equation} 
so that $e_{\tilde{q}}(0)=1$, for all $\tilde{q}$ and $ \left[ e_{\tilde{q}}(x)  \right]^a = e_{\tilde{q}/a}(ax)$, where $a$ is a constant. 
  
\section{Growth Models}
\label{sec:mathus_verhulst_gompertz}

The one-species growth models can be characterized by the \textit{saturation function}: 
\begin{equation}
G(N) = \frac{d \ln [N(\tau)]}{d\tau} \; ,
\end{equation} 
which is the \textit{per capita} growth rate. 
Time is measured as the inverse of the growth rate $1/\kappa$, i.e., $\tau=\kappa t$. 

Below, we briefly introduce the basic one-species population dynamics (growth) models. 
These models fall into two categories: one with infinite (described in terms of the number of individuals $N$) and the other with infinite ($p=N/K$) carrying capacity.
 
\subsection{Infinite Carrying Capacity}
\label{subsection-infinite}

We start presenting the Malthus model and a non-linear generalization given by the von Foerster's et al. model~\cite{vonfoerster,strazalka:2009}. 
They are described in terms of the number of individuals $N$ since they have infinite carrying capacities.

\subsubsection{Malthus Model}
\label{malthus_model}

For the Malthus model $G(N) = 1$
\begin{equation}
\frac{d N(\tau)}{d\tau} = N(\tau),
\label{eq:malthus}
\end{equation}
the result is a populational exponential growth: $N(\tau) = N_0 e^{\tau}$. 
The scaling function is: 
\begin{equation}
\tilde{s}_0(\tau) = N(\tau)
\label{eq:malthus_potential}
\end{equation}
writing 
\begin{equation}
y(\tau)= \frac{\tilde{s}_0(\tau)}{\tilde{s}_0(0)} =e^{\tau} \; ,
\label{exponential}
\end{equation}
one obtains the universal equation, that is independent from the parameter $\kappa$ (growth rate) and initial condition $N_0$.  

\subsubsection{von Foerster et al. Model}
\label{vonFoerster-model}

\begin{figure}[htbp]
\begin{center}
\includegraphics[width=\columnwidth]{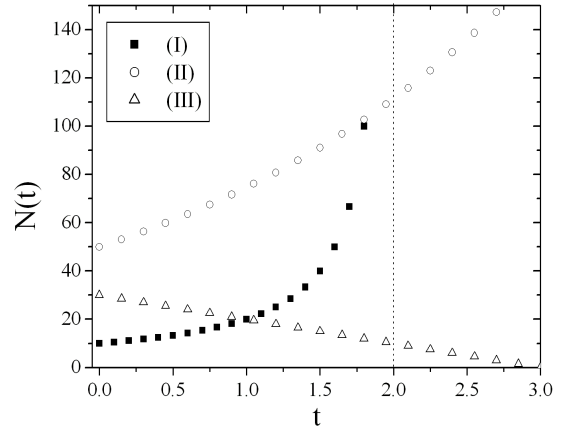} \\ {\bf (a)} \\
\includegraphics[width=\columnwidth]{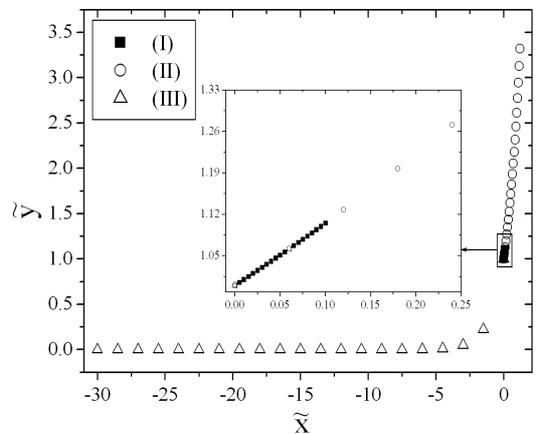} \\ {\bf (b)}
\end{center}
\caption{
{\bf (a)}: Evolution of the population according to the von Foesters model (Eq.~\ref{eq:solu_foerster}) for three different set of parameters: {\bf (I)} $N_0=10$; $\alpha=1.0 $; $\kappa=0.05$, note that the divergence occurs for $t=2.0$ {\bf (II)} $N_0=50$; $\alpha=0.0$  (Malthus);  $\kappa=0.4$, {\bf(III)} $N_0=30$; $\alpha=-1.0$; $\kappa=-10.0$. {\bf (b)}: The universal curve $y = \tilde{s}_{\alpha}(\tau)/s_F(0) = e^{\tau}$, with $\tilde{s}_{\alpha}$ given by Eq.~\ref{potential-vF}. {\bf Inset:} Zoom in of the specified region.   
}
\label{foester-fig}
\end{figure}

In the von Foerster's et al. model~\cite{vonfoerster}: $G_{\alpha}(N) = N^{\alpha}$ and: 
\begin{equation}
\frac{d  N(\tau)}{d \tau} = N(\tau)^{1 + \alpha}
\label{eq_foerster_1}
\end{equation} 
where $\alpha$ is the generalization parameter. 
Its solution is:
\begin{equation}
N(\tau) = \frac{1}{[\alpha  (T - \tau)]^{1/\alpha}} \; , 
\label{eq_sol_foerster_1}
\end{equation} 
where $T$ is a dimensionless time at which the population size diverges. 
As $\alpha \to 0$, one retrieves the Malthus model: $N(\tau) = N_0 e^{\tau}$, with $N_0 = e^{T}$. 
For human population, $\alpha \approx 2$ and a doomsday is predicted to occur on a Friday, November 13$^{\mbox{th}}$, 2026~\cite{vonfoerster}.
In terms of the one-parameter of the generalized exponential function, Eq.~(\ref{eq_sol_foerster_1}) can be written as~\cite{strazalka:2009}:
\begin{equation}
\frac{N(\tau)}{N_0} = \frac{1}{e_{\alpha}(- N_0^{\alpha} \tau)} = e_{-\alpha}( N_0^{\alpha} \tau) \; ,
\label{eq:solu_foerster}
\end{equation}
where $N_0 = N(0) = 1/(\alpha T)^{1/\alpha}$. 
Figure~(\ref{foester-fig})(a) depicts $N(\tau)$ for three different set of parameters. 

Consider Eq.~\ref{eq:solu_foerster} as $\ln_{-\alpha}[N(\tau)/N_0] = N_0^{\alpha} \tau$,  using Eq.~\ref{eq-ln-q}, one obtains: $N^{-\alpha}_0 - N^{-\alpha}(\tau) = -\alpha \tau$, after regrouping it, one writes $\ln_{-\alpha}[N(\tau)] - \ln_{-\alpha}[N_0] = \tau$  or $e^{\ln_{-\alpha}[N(\tau)]}  / e^{\ln_{-\alpha}[N_0]} = e^{\tau}$. 
In this way, we are able to define the scaling function as: 
\begin{equation}
\tilde{s}_{\alpha}(\tau) = e^{\ln_{-\alpha}[N(\tau)]} \; ,
\label{potential-vF}
\end{equation}
so that one retrieves the Malthus scaling function (Eq.~\ref{eq:malthus_potential}) for $\alpha = 0$.
Writing $y = \tilde{s}_{\alpha}(\tau)/\tilde{s}_{\alpha}(0) = e^{\tau}$, one sees that it becomes independent from the parameters and from the initial condition, presenting universality. 
This universality is depicted in Figure~(\ref{foester-fig})(b).

\subsection{Finite Carrying Capacity}
\label{subsection-finite}

The environment with limited resources is taken into account in the dependent variable $p(\tau) = N(\tau)/K$. 
Optimum environment exploration is achieved when $p(\tau) = 1$. 
Given the initial condition $\tau=\tau_0 = 0$ and $p_0 \equiv p(0)$.
The steady state solutions, according to their stability, [$d p(\tau)/d\tau|_{p^*} = 0 = p^* G(p^*)$] are either species extinction $p^* = 0$ or  survival $p^*$, obtained as the roots of $G(p^*) = 0$.  

In the following, we present the Gompertz and Verhulst models and their generalization, the Richards' model~\cite{richards_1959}. 
One way to include the finite carrying capacity $K$ to the von Foerster et al. model is to replace the number of individuals $N$ by $\ln(N/K)$ in the saturation function, this leads to the hyper-Gompertz model~\cite{turner_1976,Tsoularis:2001p2981}. 
These models have closed analytical solutions and we can collapse the solutions on a single exponential equation.  
The Tsoularis-Wallace model~\cite{tsoularis_2002} generalizes all these models, but it does not have a general analytical closed solution. 
However, we are able to write a scaling function and  show that its solution is universal.

\subsubsection{Gompertz Model}
\label{gompertz_model}

The Gompertz model, $G_{0}(p)= -\ln p$ has the solution: $p(\tau)=e^{(\ln p_0)e^{-\tau}}$. 
Writing $\ln p(\tau)/\ln p_0= e^{-\tau}$, one obtains the scaling function: 
\begin{equation}
s_{0}(\tau) = \left\{ \ln[p(\tau)] \right\} ^{-1}.
\label{eq_potencial_g}
\end{equation}
Defining $y = s_{0}(\tau)/ s_{0}(0) = e^{\tau}$, one retrieves the universal Malthusian curve.

\subsubsection{Verhulst Model}
\label{verhulst_model}

The Verhulst model $G_{1}(p) = 1-p$ has the following solution:  
\begin{equation}
p(\tau) = \frac{1}{1- (1 - p_0^{-1})e^{- \tau}} \; .
\label{verhulst-solu}
\end{equation}  
Since $\tau = \kappa t$, the model is independent from the parameter $\kappa$ but still depends on the initial condition $p_0$. 
Frequently, authors assign a partial universal behavior to the Verhulst equation: it is independent only on the parameter $\kappa$, but not on initial condition $p_0$. 
Nevertheless, calling the scaling function as 
\begin{equation}
s_1(\tau) =  [1 - p^{-1}(\tau)]^{-1}\; ,
\label{eq_potencial_v}
\end{equation}
one can write the ratio of the scaling function with respect to its initial values
\begin{equation}
y = \frac{s_1(\tau)}{s_1(0)} = \frac{[1 - p^{-1}(\tau)]^{-1}}{[1 - p_0^{-1}]^{-1}}  = e^{\tau} \; 
\end{equation}
With this procedure, one gets rid of the dependence on the initial condition and retrieves full universality. 
E. W. Montroll used this transformation in Ref~\cite[p. 4634]{Montroll_pnas_1978}  but he has not called the attention to its universal aspect.

\subsubsection{Richards' Model}
\label{section-richards-model}

Now consider the Richard's model $G_{\tilde{q}} (p) = - (p^{ \tilde{q}} -1)/\tilde{q}$~\cite{richards_1959,Montroll_pnas_1978,Strazalka:2008,martinez}, which can conveniently be written in terms $\tilde{q}$-logarithmic function
(Eq.~\ref{eq-ln-q}):   
\begin{equation} 
\frac{d \ln p(\tau)}{d \tau}  =  - \ln_{\tilde{q}} \left[ p(\tau) \right] \; .
\label{richards-model}
\end{equation}

In Ref.~\cite{Mombach:2002p1028}, the authors assume that the replication rate of a cell is regulated by a competition between the cell impetus to proliferate and an inhibition from the other cells through diffusive growth factors. 
These assumptions result in a differential equation for the growth of the cellular system identical to Eq.\ref{richards-model}, where the parameter $\tilde{q}$ is related to the range of interaction between cells and the fractal dimension where these cells grow~\cite{martinez}.
 
Notice the analogy  of Eq.~\ref{richards-model} to the Gompertz model, where we have changed the natural logarithmic function in the saturation function by the generalized one.
When discretized, this equation leads to the generalized logistic map~\cite{martinez2}. 
For $\tilde{q} = 0$, one retrieves the Gompertz model and for $\tilde{q} = 1$,  the Verhulst one.  
The solution of Eq.~(\ref{richards-model}) can be written in terms of the generalized exponential and logarithm functions as   
\begin{equation}
p(\tau)= \frac{1}{e_{\tilde{q}}\left[\ln_{\tilde{q}} \left( p_0^{-1} \right) e^{-\tau} \right]  } \;. 
\label{solution-richards-model}
\end{equation}
The asymptotic limit ($\tau \to \infty$) of  Eq.~(\ref{solution-richards-model}) is $p^* = p(\infty) = 1$, regardless the choice of $\tilde{q}$.

Let us now analyze the universality of the Richards' model. From Eq.~(\ref{solution-richards-model}) and using the properties of the generalized functions, one can write  $p(\tau)= e_{-\tilde{q}} [- \ln_{\tilde{q}} \left(p_0^{-1} \right) e^{-\tau} ]  = e_{-\tilde{q}} [ \ln_{-\tilde{q}} \left(p_0 \right) e^{-\tau}]$.
Nevertheless, taking the $-\tilde{q}$-logarithm of the preceding equation, one obtains  $\ln_{\tilde{q}}[p(\tau)] / \ln_{\tilde{q}}[p_0] = e^{-\tau}$. 
Writing the scaling function as
\begin{equation}
s_{\tilde{q}}(\tau) = \left\{\ln_{-\tilde{q}}[p(\tau)]\right\}^{-1} \; ,
\label{eq_potencial_r}
\end{equation}
one generalizes the Gompertz ($\tilde{q} = 0$, Eq.~\ref{eq_potencial_g}) and Verhulst ($\tilde{q} = 1$, Eq.~\ref{eq_potencial_v}) scaling functions. 
Defining $y = s_{\tilde{q}}(\tau)/s_{\tilde{q}}(0) =  e^{\tau}$, the Richards model presents a universal behavior, since the model is independent from the parameters ($\tilde{q}$, $\kappa$ and $K$) and from the initial condition ($p_0$). 

\subsubsection{The Hyper-Gompertz Model}
\label{section-hypergompertz-model}

The hyper-Gompertz model is obtained from the von Foerster et al. model, replacing  $N^{\alpha}$ by $[- \ln(N/K)]^{\gamma}$ and calling $p = N/K$, the saturation function becomes $G_\gamma (p) = [- \ln p]^{ \gamma}$~\cite{martinez2}, so that~\cite{turner_1976,Tsoularis:2001p2981}:
\begin{equation}
\frac{d \ln p(\tau)}{d \tau} = \{ - \ln[p(\tau)]\}^{\gamma} \; ,
\label{eq_hypergompertz}
\end{equation} 
which solution is:
\begin{equation}
\ln p(\tau) = - [(\gamma-1) \tau  -  (-1)^{\gamma} (\ln p_0)^{1 - \gamma} ]^{1/(1 - \gamma)} \; ,
\label{eq_sol_hypercompertz}
\end{equation} 
which can be written as:
\begin{equation}
[- \ln p(\tau)]^{1 - \gamma} - [- \ln p_0]^{1 - \gamma} = (\gamma-1) \tau \; ,
\end{equation}
calling the scaling function as:
\begin{equation}
s_{\gamma}(\tau) = e^{[- \ln p(\tau)]^{1 - \gamma}/(\gamma-1 )} \; ,
\end{equation}
the ratio $y =s_{\gamma}(\tau) / s_{\gamma}(0) = e^{\tau} $ is universal.
 
\subsubsection{Tsoularis-Wallace Model}
\label{sec:T-W-model}

So far, all the presented models have closed analytical solutions. 
Let us now unify these models, which presents an analytical solution, but not a general closed form. 
In terms of the $\tilde{q}$-logarithm function, the Tsoularis-Wallace model, $ G_{\alpha,\tilde{q},\gamma}(p) = p^{\alpha} [- \ln_{\tilde{q}}(p)]^{\gamma}$, is~\cite{tsoularis_2002}: 
\begin{equation}
\frac{d \ln p}{d \tau} =  p^{\alpha}(\tau) \{- \ln_{\tilde{q}}[p(\tau)] \}^{\gamma} \; . 
\label{T-W-model}
\end{equation}
The Richards' model [Eq.~\ref{richards-model}] is retrieved  for $\alpha = 0$ and $\gamma = 1$. 
For $\alpha = \tilde{q} = 0$, one retrieves the hyper-Gompertz model [Eq.~\ref{eq_hypergompertz}]. 
Writing $p = N/K$, $\gamma = 0$ and rescaling the growth rate, one retrieves von Foerster et al. model [Eq.~\ref{eq_foerster_1}]. 

The solution $p(\tau)$ of Eq.~\ref{T-W-model}  is the root of:  
\begin{equation}
B_{p^{\tilde{q}}(\tau)} \left( -\frac{\alpha}{\tilde{q}} , 1- \gamma \right)  -   B_{p_{0}^{\tilde{q}}} \left( -\frac{\alpha}{\tilde{q}} , 1- \gamma \right) = \tilde{q}^{1-\gamma}\tau  \; , 
\end{equation}
where $B_x (a,b) = \int_0^x t^{a-1}(1-t)^{b-1}dt$ is the incomplete beta function.  
Writing the scaling function as:  
\begin{equation}
 s_{\alpha,\tilde{q},\gamma}(\tau) = e^{B_{p^{\tilde{q}}(\tau)} \left( -\alpha/\tilde{q} , 1- \gamma \right)/\tilde{q}^{1-\gamma}} 
\label{potential-T-W}
\end{equation}
for $y =s_{\alpha,\tilde{q},\gamma}(\tau) / s_{\alpha,\tilde{q},\gamma}(0) = e^{\tau} $, one obtain the universality of the Tsoularis-Wallace model. 
This model illustrates that to find the scaling function, one does not imperatively has to know the analytical closed solution for $p(\tau)$. 

\section{Extinction and Survival Phases: Effort Rate}
\label{sec:generalized_model_effort_rate}

The effort rate $\tilde{\epsilon}$ quantifies insertion and removal of individuals in a population. 
Here, we investigate the effect of $\tilde{\epsilon}$ in the universality of Richards (closed analytical solution) and Tousalis-Wallace (analytical solution) models. 
Further, we point out the transition between the extinction and survival phases at a determined critical value. 
Although we do not consider the stochastic models, the time dependent effort rate can be considered as a random variable $\tilde{\epsilon}(t)$. 
In this case, one is able to deal with additive noise. 
To treat multiplicative noise, one must consider a time dependent growth rate $\kappa(t)$. 

\subsection{Constant Effort Rate}
\label{sec:constant_effort_rate}

With constant effort rate, one can deal with the transition between the extinction and survival phases. 
The steady-state solution represents the order parameter since it vanishes in the extinction phase and, at a critical point, it gives rise to a survival phase. 
This transition is present in the Richards and Tsoularis-Wallace models. 

\subsubsection{Richards-Schaefer's Model}
\label{sec:richards_schaefer_model}

\begin{figure}[htbp]
\begin{center}
		\includegraphics[width=\columnwidth]{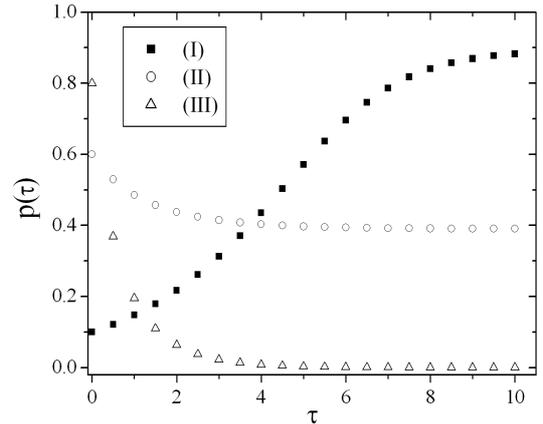} \\ {\bf (a)} \\
		\includegraphics[width=\columnwidth]{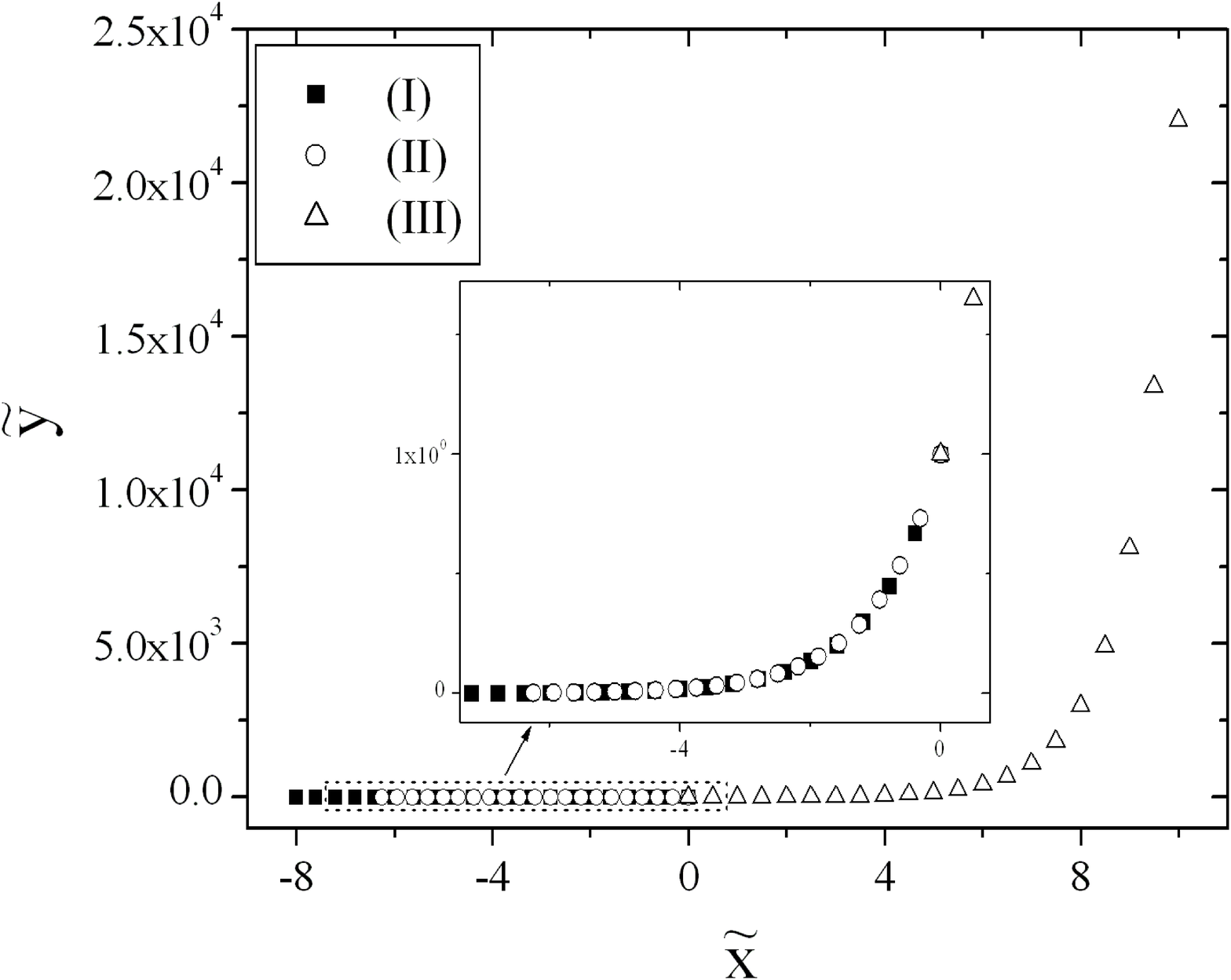} \\ {\bf (b)} \\
\end{center}
	      \caption{
\textbf{(a)}: Evolution of the population according to the Richards-Schaefer's model (Eq.~\ref{sol_richards-shaefer_eq}) as a function of $\tau$ for three different set of parameters: \textbf{(I)} $p_0=0.1;\tilde{q}=2.0;\epsilon=-0.1$, \textbf{(II)} $p_0=0.6;\tilde{q}=0.5;\epsilon=-0.75$,\textbf{(III)} $p_0=0.8;\tilde{q}=1.0;\epsilon=-2.0$.\textbf{(b)}: the universal curve $y = s_{\tilde{q}}(\tau)/s_{\tilde{q}}(0) = e^{\tau}$, with $s_{\tilde{q}}$ given by Eq.~\ref{eq_potencial_rs}. \textbf{Inset}: Zoom in of the specified region.   
}
	\label{fig:ptau-r-s}
\end{figure}

We refer to the Richards' model~(\ref{richards-model}) with a constant effort rate as the Richards-Schaefer's model: 
\begin{equation}
\frac{d \ln p(\tau)}{d \tau}=  - \ln_{\tilde{q}}[ p(\tau)]  + \epsilon \; ,
\label{mon22}
\end{equation}
where $\epsilon = \tilde{\epsilon}/\kappa$. 
The solution of the Eq.~(\ref{mon22}) is:
\begin{equation}
p(\tau)=\frac{e_{\tilde q}(\epsilon)}{e_{\tilde q}\left\{ \ln_{\tilde q}\left[\frac{e_{\tilde q}(\epsilon)}{p_0}\right] e^{-(1 + \tilde{q} \epsilon ) \tau} \right\}},
\label{sol_richards-shaefer_eq}
\end{equation}
where $p_0 = p(0)$ is the initial condition.
In Fig.~(\ref{fig:ptau-r-s}), the different behaviors of $p(\tau)$ for different set of parameters are depicted. 
For $\epsilon = 0$ in ~Eq.~(\ref{sol_richards-shaefer_eq}), one retrieves  the Richards' model Eq.~\ref{solution-richards-model}. 

\begin{figure}[htbp]
\label{equiq}
	\centering
		\includegraphics[width=\columnwidth]{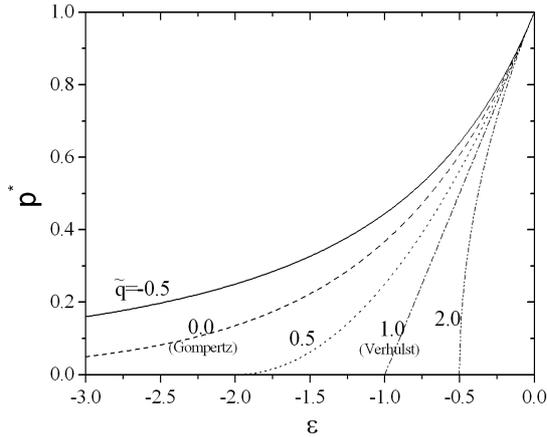}
		\caption{\label{fig-asymptotic} Asymptotic behavior of the Richards-Schaefer's Model solution, given by Eq.~(\ref{sol_richards-shaefer_eq_asymptotic}), as a function of $\epsilon$ for several values of  $\tilde{q}$. 
The species extinction ($p^* = 0 $) happen only when $\tilde{q}>0$. The critical $\epsilon$ values, i.e. the $\epsilon$ value which extinction take place, is given by Eq.\ref{eq_critical_value} }.
\end{figure}

The asymptotic behavior of Eq.~(\ref{sol_richards-shaefer_eq}) is:
\begin{equation}
p^* = p(\infty) = e_{\tilde q}(\epsilon) \; ,
\label{sol_richards-shaefer_eq_asymptotic}
\end{equation}
so that  for $\epsilon = 0$, one retrieves $p^* = 1$, as expected from Eq.~\ref{sol_richards-shaefer_eq}. 
It is interesting to point out that Eq.~(\ref{sol_richards-shaefer_eq_asymptotic}) vanishes for $\tilde{q} \epsilon \le -1$, representing species extinction (see Fig.~\ref{fig-asymptotic}).  
Species survival occurs for $\epsilon > \epsilon^{(c)}$, where the critical value is:  
\begin{equation}
\epsilon^{(c)} = -\frac{1}{\tilde{q}} \; .
\label{eq_critical_value}
\end{equation}
One thinks of $p^*$ as an order parameter that describes two ecological stable phases: species extinction and survival. 
These phases are separated by the critical value $\epsilon_c$. 
This transition is supressed for $\tilde{q} = 0$, the Gompertz model.

To show the universality of this model, write  $y(\tau)=p(\tau)/e_{\tilde q}(\epsilon)$, so that Eq.~(\ref{sol_richards-shaefer_eq})  becomes $y(\tau) = e_{-\tilde q} \{ \{\ln_{-\tilde q}(y_0) \exp\{[ -(1+ \tilde{q}\epsilon)\tau \} \}$, where $y_0 = y(0)$ and  we have used the properties:   $1/e_{\tilde q}(x)=e_{-\tilde q}(-x)$ and $-\ln_{\tilde q}(x^{-1})=\ln_{-\tilde q}(x)$. 
Applying $- \tilde{q}$-logarithm on $y(\tau)$ one has: $\ln_{-\tilde q}[y(\tau)]=  \ln_{- \tilde q}(y_0) e_{\tilde q}(\epsilon) \exp\left\{[ -(1+\tilde{q} \epsilon)    \tau \right\}$.
Thus Eq.~\ref{sol_richards-shaefer_eq} can be written as 
$\ln_{-\tilde{q}}[ p(\tau)/e_{\tilde q}(\epsilon)] / \ln_{-\tilde{q}}[ p_0/e_{\tilde q}(\epsilon)] = e^{-(1 + \tilde{q} \epsilon) \tau}$. The scaling function is given by: 
\begin{equation}
s_{\tilde{q},\epsilon}(\tau) =  \left\{ \frac{\ln_{-\tilde{q}}\left[p(\tau)/e_{\tilde{q}(\epsilon)} \right]}{e_{\tilde{q}}(\epsilon)}\right\}^{-\left[e_{\tilde{q}}(\epsilon)\right]^{-\tilde{q}}} \;. 
\label{eq_potencial_rs}
\end{equation}
So that $\epsilon = 0 \Rightarrow e_{\tilde q}(0) = 1$ and one retrieves Eq.~\ref{eq_potencial_r}. 
Calling $y= s_{\tilde{q}}(\tau) / s_{\tilde{q}}(0) = e^{\tau}$, one retrieves the universal Malthus model.
The introduction of $e_{\tilde{q}}(\epsilon)$ in the denominator of the scaling function, which does not affect the result, is due the necessity of it to be compatible with the time dependent effort rate. 

\subsubsection{Tsoularis-Wallace-Schaefer Model}
\label{tsoularis_wallace_schaefer_model}

Let us consider an effort rate in the Tsoularis-Wallace model, and call it the Tsoularis-Wallace-Schaefer model: 
\begin{equation}
 \frac{d \ln p(\tau)} {d \tau}  =    p^{\alpha}(\tau) \left\{ - \ln_{\tilde{q}} [p(\tau)]  \right\}^{\gamma}  + \epsilon
\label{eq:tsoularis-wallace-schaefer}
\end{equation}
which solution $p(\tau)$ is the root of 
\begin{equation}
\int_{p_0}^{p(\tau)} \frac{dx}{ x \{ x^{\alpha}[- \ln_{\tilde{q}}(x)]^{\gamma} + \epsilon \} }  =  \tau
\end{equation}

The scaling function is defined as
\begin{equation}
s_{\alpha,\tilde{q},\gamma,\epsilon}(\tau) = \exp \left[ \int_{0}^{p(\tau)} \frac{dx}{ x \{  x^{\alpha}[- \ln_{\tilde{q}}(x)]^{\gamma} + \epsilon \}  } \right]
\end{equation}
so that a universal behavior is found for $y = s_{\alpha,\tilde{q},\gamma,\epsilon}(\tau)/s_{\alpha,\tilde{q},\gamma,\epsilon}(0) = e^{\tau}$, even so we have not been able to calculate the integral to find explicitly $p(\tau)$. 

The steady-state solution is obtained considering $d p/d \tau = 0$ in Eq.~\ref{eq:tsoularis-wallace-schaefer}, so that $p^{*} \{ p^{* \alpha} [- \ln_{\tilde{q}}(p^*)  ]^{\gamma}  + \epsilon \} = 0$, leading to $p^* = 0$, which represents the extinction phase and $p^{* \, \alpha} [- \ln_{\tilde{q}}(p^*)  ]^{\gamma}  = - \epsilon$, which represents the survival phase. 
The critical value separationg the extinction and survival phase, is given by the root of: $(p^{*})^{\alpha/\gamma} - (p^{*})^{\alpha/\gamma + \tilde{q}} = \tilde{q}(-\epsilon)^{1/\gamma}$.

\begin{figure*}[htbp]
		\includegraphics[width=1.7\columnwidth]{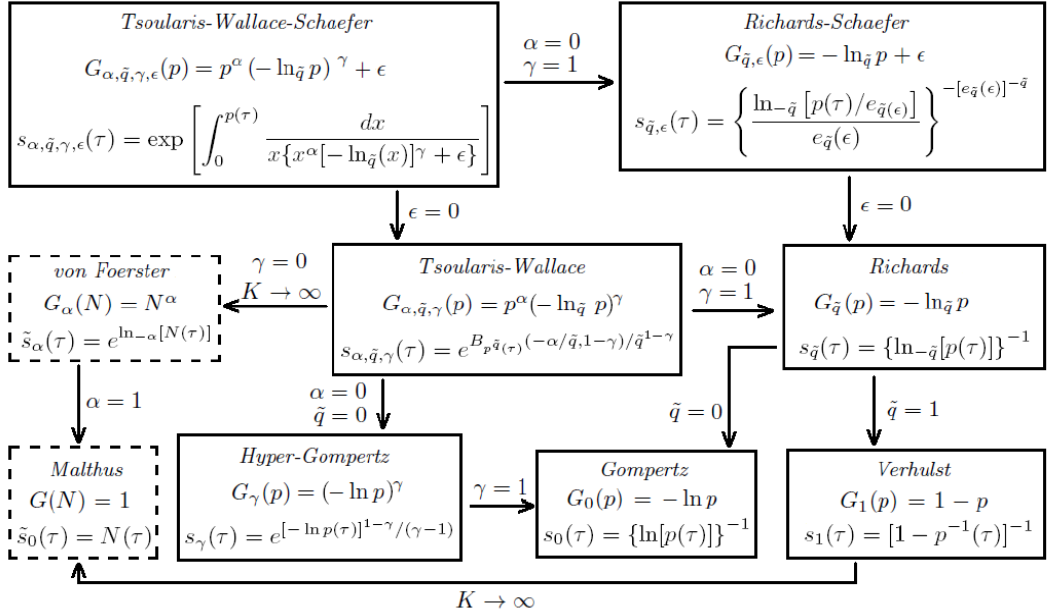}
	      \caption{The considered growth models from sections \ref{sec:mathus_verhulst_gompertz} and \ref{sec:constant_effort_rate} with their respective quantities: Saturation function $G$ and scaling function $s$.
	      The Malthusian universal behavior (independence of parameters and initial conditions) is obtained in all models by the ratio of the scaling function in relation to its initial value:  $y = s(\tau)/ s(0) = e^{\tau}$.}
	\label{fig:bigfig}
\end{figure*}

\subsection{Time Dependent Effort Rate}
\label{time_dependent_effort_rate}

It is interesting to consider time dependent effort rate once it can be considered as a random variable and noise can be treated. 
On one hand, if the growth rate is constant, the noise is additive to the model. 
On the other hand, if the growth rate is time dependent and a random variable, one can consider the effort rate to vanish and one has multiplicative noise in the system.

Although we do not address the stochastic models, we present the analytical solutions of these models and show their universalities. 

\subsubsection{Constant Growth Rate}
\label{constant_growth_rate}

Consider now a time dependent \textit{effort rate} $\tilde{\epsilon}(t)$ in the Richards-Schaefer's model [Eq.~(\ref{mon22})]:
\begin{equation}
\frac{d \ln p(\tau)}{d \tau}= -  \ln_{\tilde q}p(\tau) + \epsilon(\tau) \; .
\label{time-dependent-model2}
\end{equation} 
The solution of Eq.~(\ref{time-dependent-model2}) is 
\begin{eqnarray}
p(\tau) & = & \frac{e_{\tilde q}\left[ \epsilon(\tau) \right]}{e_{\tilde q}\left\{\ln_{\tilde q}\left\{\frac{e_{\tilde q}\left[\epsilon(0)\right]}{p_0}\right\} \frac{e_{\tilde q}[\epsilon(\tau)]}{e_{\tilde q}[\epsilon(0)]} e^{-\left[1 + \tilde{q} \overline{\epsilon}(\tau) \right] \tau}\right\}} 
\label{eq:sol_time_richards_shaefer_model} 
\end{eqnarray}
where 
\begin{eqnarray}
\overline{\epsilon}(\tau) & = & \frac{1}{\tau} \int_0^{\tau} d\tau' \epsilon(\tau') 
\label{eq:media_alpha} \; . 
\end{eqnarray}
is the mean value of $\epsilon(\tau)$ up to time $\tau$. 
For a constant effort rate $\epsilon(\tau)=\epsilon$ in Eq.~\ref{eq:sol_time_richards_shaefer_model}, one retrieves 
the Richards-Schaefer's model and its solution [Eq.~(\ref{sol_richards-shaefer_eq})]. 

The steady state solution ($\tau \to \infty$) of Eq.~\ref{eq:sol_time_richards_shaefer_model} is:  
\begin{equation}
p^* = p(\infty) = e_{\tilde{q}}(\overline{\epsilon}) \; ,
\end{equation}
where $\overline{\epsilon} = \overline{\epsilon(\infty)}$ is the true mean value of $\overline{\epsilon(\tau)}$. 
Species extinction occurs for $\tilde{q} \overline{\epsilon}  < -1$.

Write  $y(\tau)=p(\tau)/e_{\tilde q}[\epsilon(\tau)]$, so that Eq.~(\ref{eq:sol_time_richards_shaefer_model})  becomes $y(\tau) = e_{-\tilde q} \{ \{\ln_{-\tilde q}(y_0)/e_{\tilde q}[\epsilon(0)]\} $ $ e_{\tilde q}[\epsilon(\tau)]$ $ \exp\{-[1 + \tilde{q} \overline{\epsilon}(\tau) ] \tau \} \}$, where $y_0 = y(0)$ and we have used:   $1/e_{\tilde q}(x)=e_{-\tilde q}(-x)$ and $-\ln_{\tilde q}(x^{-1})=\ln_{-\tilde q}(x)$. 
Applying $- \tilde{q}$-logarithm on $y(\tau)$ one has: $\ln_{-\tilde q}[y(\tau)]=  \ln_{- \tilde q}(y_0) e_{\tilde q}[\epsilon(\tau)]/e_{\tilde q}[\epsilon(0)] \exp\left\{-[1 + \tilde{q} \overline{\epsilon}(\tau)] \tau \right\}$, justifing the definition of the scaling function as 
\begin{equation}
s_{\tilde{q},\epsilon(\tau)}(\tau) = \left\{ \frac{\ln_{- \tilde q}\left[\frac{p(\tau)} {e_{\tilde q}[\epsilon(\tau)]}\right]}{e_{\tilde q}[\epsilon(\tau)]} \right\}^{-e_{\tilde{q}}[\epsilon(\tau)]^{- \tilde{q}}} \; .
\end{equation}
Taking $\epsilon(\tau) = \epsilon(0) = \epsilon$, one retrieves Eq.~ \ref{eq_potencial_rs}. 
Notice that it becomes clear the reason we have introduced the factor $1/e_{\tilde q}(\epsilon)$ in the definition of the potential growth 
on the Richard-Schaefer model with constant effort rate Eq.~\ref{eq_potencial_rs}. 

Defining $y = s_{\tilde{q},\epsilon(\tau)}(\tau)/s_{\tilde{q},\epsilon(\tau)}(0)  = e^{\tau}$, one obtains the universal Malthusian equation, independent of parameters and initial conditions.

If $\epsilon(t)$ is a random variable, then one has the additive stochastic growth equation. 
In this case, if its mean value vanishes $\overline{\epsilon(\tau)} = 0$ and $\overline{\epsilon(\tau_1)\epsilon(\tau_2)} = \sigma^2 \delta(\tau_2 - \tau_1)$ (Gaussian process), then the probability density function  of $v  = \ln p$ satisfies the Fokker-Planck equation: $\partial_{\tau} P(v) =  \partial_{v}[P (v)  \ln_{\tilde q}(v)] + (\sigma^2/2) \partial^2_{v}[P(v)]$~\cite{Montroll_pnas_1978,zygadlo_1993a,zygadlo_1993b}.
Correlated and L\'evy like noise have also been addessed~\cite{mannella_1990,ai_2003,spagnolo_2008}. 

\subsubsection{Time Dependent Growth Rate }
\label{sec_full-richards_shaeffer_model}

Consider now time dependence in both growth and effort rates.
One has:  
\begin{equation}
\frac{d \ln [p(t)]}{d t}= - \kappa(t) \ln_{\tilde q}[p(t)] + \tilde{\epsilon}(t).
\label{dlnp}
\end{equation}
Notice that here we use $t$ instead of $\tau$ as the independent variable.
The soultion of Eq. \ref{dlnp} is given by:
\begin{equation}
p(t)=\left\{\frac{1}{\tilde{I}(t)}\left[\int_0^t dt' \tilde{I}(t')\kappa(t')+p_0^{-\tilde{q}}\right]\right\}^{-1/\tilde{q}}
\label{solp}
\end{equation}
where:
\begin{equation}
\tilde{I}(t)=e^{\left[\int_0^{t} dt'\kappa(t')+\tilde{q}\int_0^{t} dt' \tilde{\epsilon}(t')\right]}
\label{it}
\end{equation}
so that $I(0) = 1$.

\begin{equation}
\tilde{I}(t) p^{-\tilde{q}}(t) - \tilde{I}(0) p_0^{-\tilde{q}} = \int_0^t dt'I(t')\kappa(t')
\end{equation}
so that 
\begin{equation}
\frac{e^{\tilde{I}(t) p^{-\tilde{q}}(t)} }{e^{\tilde{I}(0) p_0^{-\tilde{q}}}} = e^{\int_0^t dt'I(t')\kappa(t')} \; .
\end{equation}
Calling $\tilde{\tau} = \int_0^t dt'I(t')\kappa(t')$ and the scaling function as:
\begin{equation}
s_{\kappa(t),\tilde{q},\tilde{\epsilon}(t)}(t) = e^{\tilde{I}(t) p^{-\tilde{q}}(t)}  \; .  
\end{equation}
The ratio $y = s_{\kappa(t),\tilde{q},\tilde{\epsilon}(t)}(t)/s_{\kappa(t),\tilde{q},\tilde{\epsilon}(t)}(0) = e^{\tau}$ is universal. 

If $\tilde{\epsilon}(t) = 0$, one can now consider the time dependend growing rate as $\kappa(t) = a_0(t) + a_1 \gamma_1(t)$, where $a_0(t)$ a determinist growth and $\gamma_1(t)$ may be considered as a multiplicative stochastic noise~\cite{calisto_2007,aquino_2010}.

\section{Conclusion}
\label{conclusion}

We show that measuring time $\tau = t/\kappa$, where $\kappa$ is the growth rate, we are able to write a general expression to obatain  the \textit{scaling function}:  
\begin{equation}
s(\tau)=e^{\int_0^{\tau} d v / G(e^v) } \; ,
\end{equation}
where, $v = \ln N$ and $G(N)$ is the saturation function.
Using the scaling function, all the considered models are written as the universal Malthus (exponential) model. 
We have shown the universal properties of the Tsoularis-Wallace-Schaerfer model, which is a very general constant coefficient model,  with no closed analytical solution. 
Although we have not derived all the possible particular cases from this model, we can deduce that they are all universal.
If one includes the effort rate, a transition from species extinction to the survival is well determined.
This effort rate may represent the mean field approximation of the interaction of other species. 
For this reason, we believe in the universality of multi-species models.  
For time dependent coefficients, the most general model we have addressed and solved is the Richards-Schaeffer model, which is also universal.
Since one can consider stochasticity in the time dependent coefficient models, either with additive and multiplicative noise, 
we conjecture that the stochastic models are also universal.
The universality is useful when working with experimental data as the model that best fits the data must correspond to a straight line in a data collapsed  semi-log graph.

\section*{Ackowledgements}

A.S.M. acknowledges support from CNPq (303990/2007-4). 
B. C. T. C. acknowledges support from CAPES.
F. R. acknowledges support from CNPq (151057/2009-5). 
The authors thank Dominik Strzalka for pointing out Ref.~\cite{strazalka:2009} and consequently Ref.~\cite{vonfoerster}.

\end{document}